# Simulation of a wire-cylinder-plate positive corona discharge in nitrogen gas at atmospheric pressure

**Alexandre A. Martins**
Institute for Plasmas and Nuclear Fusion & Instituto Superior Técnico,
Av. Rovisco Pais, 1049-001 Lisboa, Portugal

**Abstract:** In this work we are going to perform a simulation of a wire-cylinder-plate positive corona discharge in nitrogen gas, and compare our results with already published experimental results in air for the same structure. We have chosen to simulate this innovative geometry because it has been established experimentally that it can generate a thrust per unit electrode length transmitted to the gas of up to 0.35 N/m and is also able to induce an ion wind top velocity in the range of 8-9 m/s in air. In our model, the used ion source is a small diameter wire, which generates a positive corona discharge in nitrogen gas directed to the ground electrode, after which the generated positive ions are further accelerated in the acceleration channel between the ground and cathode. By applying the fluid dynamic and electrostatic theories all hydrodynamic and electrostatic forces that act on the considered geometries will be computed in an attempt to theoretically confirm the generated ion wind profile and also the thrust per unit electrode length. These results are important to establish the validity of this simulation tool for the future study and development of this effect for practical purposes.

## I. INTRODUCTION

The preliminary experimental results of Dorian Colas and his colleagues[1,2] on a new electrostatic wind generator have demonstrated the advantage of using direct current/voltage corona discharges in order to produce useful ion wind velocities that, like plasma actuators, could also be applied for many different practical purposes. This novel configuration may be useful to substitute small wind fans or propulsion units for small UAV's (unmanned aerial vehicles), or possibly improve the efficiency of turbo-machinery by reducing turbulence, or be used for aerodynamic boundary layer control, aircraft drag reduction or wing lift increase, noise reduction or other internal flow configurations, with the advantage of using no moving mechanical parts.[3-11]

Our present work has the purpose to perform a theoretical verification of the innovative Dorian Colas team results[1,2] regarding the thrust per unit electrode length transmitted to the gas and also the ion wind profile generated in air so that we can validate our simulation approach and also improve on our understanding of the physical principles responsible for the performance of the setup put forward by them, in the hope that we will be able to improve this system in the future.

The structure to be studied[1,2] is represented in figure 1 (the space between each dot represents 1 mm). We are going to implement a numerical simulation of this geometry in nitrogen gas (as a first approximation to air), at atmospheric pressure, in order to obtain the resultant electro-hydrodynamic (EHD) flow when a positive corona discharge is established. In this configuration we have only one corona wire centered at (-0.016 m, 0 m) and distanced 6 mm from the plane represented by the two ground spheres (center of the spheres), which are centered at (-0.01 m, -0.007 m) and (-0.01 m, 0.007 m), and that are distanced between themselves by 8 mm, as can be seen in figure 1. The cathode electrodes have 20 mm in length and 8 mm in height, being charged to -8000 Volts, and separated to the ground spheres by 4 mm (figure 1). The positive corona wire has a radius of 0.1 mm, which is much smaller than the radius of 6 mm of the ground spheres.

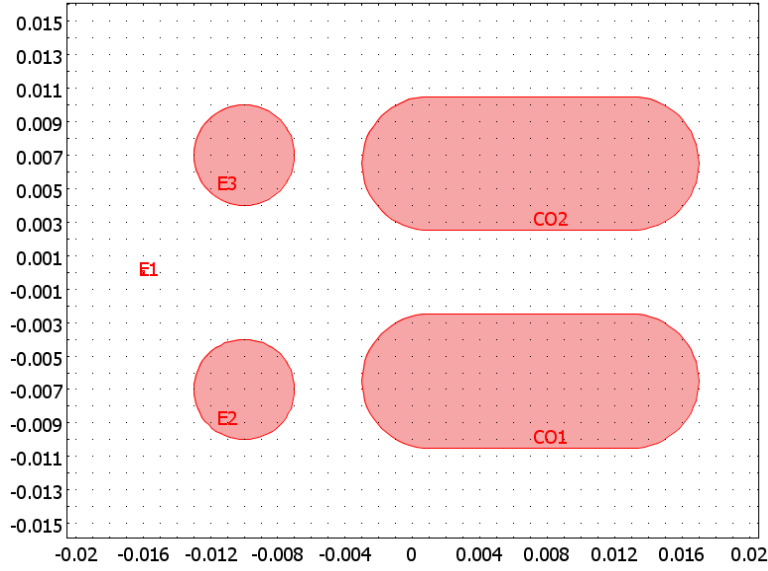

**Figure 1.** The considered geometry with the corona wire (E1) at (- 0.016 m, 0 m), two grounded spheres (E2 and E3) 6 mm in diameter and centered respectively at (-0.01 m, -0.007 m) and (-0.01 m, 0.007 m), and two cathode electrodes CO1 and CO2 (the used grid represents spaces of 1 mm).

## II. NUMERIC MODEL

In our numerical model we will consider electrostatic force interactions between the electrodes and the ion space cloud, the moment exchange between the mechanical setup and the induced opposite direction nitrogen flow (EHD flow), nitrogen pressure forces on the structure and viscous drag forces. EHD flow is the flow of neutral particles caused by the drifting of ions in an electric field. In our case these ions are generated by a positive high voltage corona discharge in the high curvature (the higher the radius of a sphere, the less is its curvature) portions of the electrodes. The corona wire has a uniform high curvature and therefore will generate ions uniformly.[12] The positively ionized gas molecules will travel from the corona wire ion source towards the collector (ground) colliding with neutral molecules in the process. These collisions will impart momentum to the neutral atoms that will move towards the collector as a result. The momentum gained by the neutral gas is exactly equal to the momentum gained by the positive ions accelerated through the electric field between the electrodes, and lost in inelastic collisions to the neutral molecules.

Corona discharges are non-equilibrium plasmas with an extremely low degree of ionization (roughly $10^{-8}$ %). There exist two zones with different properties, the ionization zone and the drift zone.[13] The energy from the electric field is mainly absorbed by electrons in the ionization zone, immediately close to the corona electrode, and transmitted to the neutral gas molecules by inelastic collisions producing electron-positive ion pairs, where the net space charge density $\rho_q$ will remain approximately zero ($\rho_q = 0$). However, the local volume space charge, in the drift zone, will be positive for a positive corona (and therefore $\rho_q = eZ_i n_i$ in the drift zone; $e$ is absolute value of the electron charge, $Z_i$ is the charge of the positive ionic species, $n_i$ is the positive ion density of $N_2^+$) because of the much higher mobility of electrons relative to the positive ions, and because the only region were (positive) ions will be generated is in the ionization region where the electrons have enough energy to accomplish that due to the much higher electric field intensity. Also, in the drift region the electrons and ions do not have enough energy to react with neutrals and also have too low density to react with other ionized particles.[13] Therefore, we will only consider nitrogen ($N_2^+$) positive ions in the drift zone, which is the most relevant nitrogen ion in atmospheric processes.[12]

The ionic mobility ($\mu_i$) is defined as the velocity v attained by an ion moving through a gas under unit electric field $E$ (m$^2$ s$^{-1}$ V$^{-1}$), i.e., it is the ratio of the ion drift velocity to the electric field strength:

$$\mu_i = \frac{v}{E}. \tag{1}$$

The mobility is usually a function of the reduced electric field $E/N$ and $T$, where $E$ is the field strength, $N$ is the Loschmidt constant (number of molecules m$^{-3}$ at s.t.p.), and $T$ is the temperature of the gas. The unit of $E/N$ is the Townsend (Td), 1 Td = $10^{-21}$ V m$^2$. Since we are applying 14000 V to the corona wire across a gap of 6.22 mm towards the ground electrode, the reduced electric field will be approximately 92 Td or 92 × $10^{-17}$ Vcm$^2$ (considering that the gas density $N$ at 1 atm, with a gas temperature $T$ of 300 K is $N$=2.447×10$^{19}$ cm$^{-3}$). According to Moseley,[14] the mobility $\mu_i$ of an ion is defined by:

$$\mu_i = \mu_{i0}(760/p)(T/273.16), \tag{2}$$

where $\mu_{i0}$ is the reduced mobility, $p$ is the gas pressure in Torr (1 atm = 760 Torr) and $T$ is the gas temperature in Kelvin. For our experimental condition of $E/N$ = 92 Td, Moseley's measurements indicate a $\mu_{i0}$ of 1.60 cm$^2$/(Vs). Thus, at our operating temperature of 300 K, the mobility $\mu_i$ will be 1.76 cm$^2$/(Vs) or 1.76×10$^{-4}$ m$^2$/(Vs).

Since the reduced electric field is relatively low, the ion diffusion coefficient $D_i$ can be approximated by the Einstein relation:

$$D_i = \mu_i \left(\frac{k_B T}{e}\right), \tag{3}$$

where $k_B$ is the Boltzmann constant. This equation provides a diffusion coefficient of 4.54×10$^{-6}$ m$^2$/s for our conditions.

The governing equations for EHD flow in an electrostatic fluid accelerator (EFA) are already known[15,16] and described next; these will be applied to the drift zone only. The electric field **E** is given by:

$$\mathbf{E} = -\nabla V. \tag{4}$$

Since $\nabla \cdot \mathbf{E} = \frac{\rho_q}{\varepsilon_0}$ (Gauss's law), the electric potential $V$ is obtained by solving the Poisson equation:

$$\nabla^2 V = -\frac{\rho_q}{\varepsilon_0} = -\frac{e(Z_i n_i - n_e)}{\varepsilon_0}, \qquad (5)$$

where $n_e$ is the electron density (we are not considering negative ions) and $\varepsilon_0$ is the permittivity of free space. The total volume ionic current density $\mathbf{J_i}$ created by the space charge drift is given by:

$$\mathbf{J}_i = \rho_q \mu_i \mathbf{E} + \rho_q \mathbf{u} - D_i \nabla \rho_q, \qquad (6)$$

where $\mu_i$ is the mobility of ions in the nitrogen gas subject to an electric field, $\mathbf{u}$ is the gas (nitrogen neutrals) velocity and $D_i$ is the ion diffusion coefficient. The current density satisfies the charge conservation (continuity) equation:

$$\frac{\partial \rho_q}{\partial t} + \nabla \cdot \mathbf{J}_i = 0. \qquad (7)$$

But, since we are studying a DC problem, in steady state conditions we have:

$$\nabla \cdot \mathbf{J}_i = 0. \qquad (8)$$

The hydrodynamic mass continuity equation for the nitrogen neutrals is given by:

$$\frac{\partial \rho_f}{\partial t} + \nabla \cdot (\rho_f \mathbf{u}) = 0. \qquad (9)$$

If the nitrogen fluid density $\rho_f$ is constant, like in incompressible fluids, then it reduces to:

$$\nabla \cdot \mathbf{u} = 0. \qquad (10)$$

In this case, the nitrogen is incompressible and it must satisfy the Navier-Stokes equation:

$$\rho_f \left( \frac{\partial \mathbf{u}}{\partial t} + (\mathbf{u} \cdot \nabla)\mathbf{u} \right) = -\nabla p + \mu \nabla^2 \mathbf{u} + \mathbf{f}. \qquad (11)$$

The term on the left is considered to be that of inertia, where the first term in brackets is the unsteady acceleration, the second term is the convective acceleration and $\rho_f$ is the density of the hydrodynamic fluid - nitrogen in our case. On the right, the first term is the pressure gradient, the second is the viscosity ($\mu$) force and the third is ascribed to any other external force $\mathbf{f}$ on the fluid. Since the discharge is DC, the electrical force density on the nitrogen ions that is transferred to the neutral gas is $\mathbf{f^{EM}} = \rho_q \mathbf{E} = -\rho_q \nabla V$. If we insert the current density definition (equation (6)) into the current continuity (equation (8)), we obtain the charge transport equation:

$$\nabla \cdot \mathbf{J}_i = \nabla \cdot (\rho_q \mu_i \mathbf{E} + \rho_q \mathbf{u} - D_i \nabla \rho_q) = 0. \qquad (12)$$

Since the fluid is incompressible ($\nabla \cdot \mathbf{u} = 0$) this reduces to:

$$\nabla \cdot (\rho_q \mu_i \mathbf{E} - D_i \nabla \rho_q) + \mathbf{u} \nabla \rho_q = 0. \tag{13}$$

In our simulation we will consider all terms present in equation (13), although it is known that the conduction term (first to the left) is preponderant over the other two (diffusion and convection), since generally the gas velocity is two orders of magnitude smaller than the velocity of ions. Usually, the expression for the current density (equation (6)) is simplified as:

$$\mathbf{J}_i = \rho_q \mu_i \mathbf{E}. \tag{14}$$

Then, if we insert equation (14) into equation (8), expand the divergence and use equation (4) and Gauss's law we obtain the following (known) equation that describes the evolution of the charge density in the drift zone:

$$\nabla \rho_q \cdot \nabla V - \frac{\rho_q^2}{\varepsilon_0} = 0. \tag{15}$$

In table I we can see the values of the parameters used for the simulation. We will consider in our model that the ionization region has zero thickness, as suggested by Morrow.[17] The following equations will be applied only to the ionization zone in order to determine the proper boundary condition that we would have in the boundary between the ionization and drift zones and apply that directly on the surface of the corona wire, so that we take into account any ionization zone effects in our model. For the formulation of the proper boundary conditions for the external surface of the space charge density we will use the Kaptsov hypothesis[18] which states that below corona initiation the electric field and ionization radius will increase in direct proportion to the applied voltage, but will be maintained at a constant value after the corona is initiated.

In our case, a positive space charge $\rho_q$ is generated by the corona wire and drifts towards the ground electrode through the gap (drift zone) between both electrodes and is accelerated by the local electric field. When the radius of the corona wire is much smaller than the gap, then the ionization zone around the corona wire is uniform. In a positive corona, Peek's empirical formula[19-22] in air gives the electric field strength $E_p$ (V/m) at the surface of an ideally smooth cylindrical wire corona electrode of radius $r_c$:

$$E_p = E_0 \cdot \delta \cdot \varepsilon (1 + 0.308 / \sqrt{\delta \cdot r_c}). \tag{16}$$

Where $E_0 = 3.31 \times 10^6 V/m$ is the dielectric breakdown strength of air (we used the nitrogen breakdown strength, which is 1.15 times higher than that for air,[20] $\delta$ is the relative atmospheric density factor, given by $\delta = 298p/T$, where $T$ is the gas temperature in Kelvin and $p$ is the gas pressure in atmospheres ($T$=300K and $p$=1atm in our model); $\varepsilon$ is the dimensionless surface roughness of the electrode ($\varepsilon = 1$ for a smooth surface) and $r_c$ is given in centimeters. At the boundary between the ionization and drifting zones the electric field strength is equal to $E_0$ according to the Kaptsov assumption. This formula (Peek's law) determines the threshold strength of the electric field to start the corona discharge at the corona wire. Surface charge density will then be calculated by specifying the applied electric potential $V$ and assuming the electric field $E_p$ at the surface of the corona wire. The assumption that the electric field strength at the wire is equal to $E_p$ is justified and discussed by Morrow.[17] Although $E_p$ remains constant after corona initiation, the space charge current

**J<sub>i</sub>** will increase with the applied potential $V_c$ in order to keep the electric field at the surface of the corona electrode at the same Peek's value, leading to the increase of the surrounding space charge density and respective radial drift.

Atten *et al*[22] have compared Peek's empirical formula with other methods including the direct Townsend ionization criterion and despite some differences in the electric field, they concluded that the total corona current differs only slightly for small corona currents (below 6 kV). For voltages above 6 kV (corresponding to higher space currents) the difference is smaller than 10% in the worst case, according to them.

For relatively low space charge density in DC coronas, the electric field *E(r)* in the plasma (ionization zone) has the form:[12]

$$E(r) = \frac{E_p r_c}{r}, \quad (17)$$

where *r* is the radial position from the center of the corona wire. Since the electric field $E_0$ establishes the frontier to the drift zone, using this formula we can calculate the radius of the ionization zone ($r_i$), which gives:

$$r_i = \frac{E_p r_c}{E_0} = r_c \cdot \delta \cdot \varepsilon (1 + 0.308/\sqrt{\delta \cdot r_c}). \quad (18)$$

Since we have chosen in our simulation for $r_c$ to be 0.1 *mm*, then $r_i$ would be 0.197 *mm*. Now we can calculate the voltage ($V_i$) at the boundary of the ionization zone by integrating the electric field between $r_c$ and $r_i$:

$$V_i = V_c - E_p r_c \ln(E_p/E_0), \quad (19)$$

where $V_c$ is the voltage applied to the corona electrode and $r_c$ is in meters. This equation is valid only for the ionization zone. In our case it determines that if we apply 14000 Volts to the corona wire, then the voltage present at the boundary of the ionization zone becomes 11816.26 Volts, which is the voltage we apply directly to the surface of the corona wire in our simulation.

For the drift zone, Poisson equation (equation (5)) should be used together with the charge transport equation (equation (13)) in order to obtain steady state field and charge density distributions. The values of the relevant parameters for the simulation are detailed in table I.

Three application modules of the COMSOL 3.5 Multiphysics software are used. The steady state incompressible Navier-Stokes mode is used to resolve the fluid dynamic equations. The electrostatics mode is used to resolve the electric potential distribution and the electrostatic forces to which the electrodes are subjected. The PDE (coefficient form) mode is used to resolve the charge transport equation (equation (13)). The parameters used for the simulation are shown in table I. The typical mesh of the solution domain consists of 50400 elements in a square of 0.2 m by 0.2 m as shown on figure 2.

Dirichlet boundary conditions were used in the PDE (coefficient form) module, where in the corona wire element an initial ion concentration of $1.2 \times 10^{-2}$ [C/m$^3$] was used, and a zero ion concentration on the ground electrodes, cathodes and on the domain frontiers. In the Incompressible Navier-Stokes module, an open boundary condition was implemented on the domain frontiers, and a wall (no slip) condition was implemented on all electrodes. In the Electrostatics module a zero charge/symmetry boundary condition was implemented on the domain frontiers, a Ground potential on the ground electrode, an electric potential of 14000 Volts on the corona wire, which according to Equation (19) transforms to 11816.26 Volts, and a cathode potential of -8000 Volts.

**Table I. Value of parameters used for the simulation.**

| Parameters | Value |
|---|---|
| Nitrogen density (T=300K, p=1atm), $\rho_N$ | 1.165 kg/m$^3$ |
| Dynamic viscosity of nitrogen (T=300K, p=1atm), $\mu_N$ | 1.775 x 10$^{-5}$ Ns/m$^2$ |
| Nitrogen relative dielectric permittivity, $\varepsilon_r$ | 1 |
| $N_2^+$ mobility coefficient, $\mu_i$ (for E/N = 92 Td) | 1.76 x 10$^{-4}$ m$^2$/(Vs) |
| $N_2^+$ diffusion coefficient, $D_i$ (for E/N = 92 Td) | 4.54 x 10$^{-6}$ m$^2$/s |
| Corona wire radius, $r_c$ | 0.1 mm |
| Ground electrode diameter | 6 mm |
| Distance between ground spheres | 8 mm |
| Distance between wire and ground | 6.22 mm |
| Cathode electrode height | 8 mm |
| Cathode electrode length | 20 mm |
| Distance between cathodes | 5 mm |
| Distance between ground and cathode | 4 mm |
| Corona wire voltage, $V_c$ | 14000 V |
| Ground electrode voltage, $V_g$ | 0 V |
| Cathode Voltage | -8000 V |

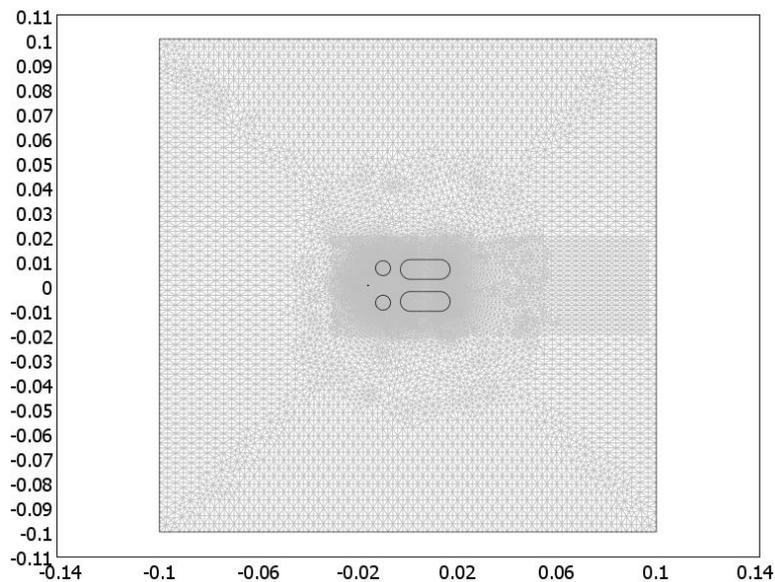

**Figure 2.** Typical mesh of the solution domain (0.2 m G 0.2 m) containing 50400 elements.

## III. NUMERICAL SIMULATION RESULTS

In previous similar simulation results using Comsol 3.5 we have demonstrated that our theoretical results have a stability or convergence up to the third decimal place above 30000 mesh elements.[23] In the current setup the corona wire generates a positive charge cloud which is accelerated through the gap towards the facing ground spheres and cathode electrodes. The interaction between all electrodes and the positive charge cloud will accelerate the nitrogen positive ions towards the grounded spheres and the ions will transmit their momentum to the neutral nitrogen particles by a collision process. Thus, the neutral nitrogen will move in the direction from the corona wire to the grounded spheres and the nitrogen neutral wind profile generated by the collisions with the accelerated ion cloud is shown in figure 3, were the top velocity achieved is 8.448 m/s. The electric field vector distribution around the asymmetric capacitor (without the ion space charge) is shown in figure 4, and the ion density profile is represented in figure 5 from (-0.05 m, 0) to (0.05 m, 0). The electrostatic force vectors for the electrodes in this configuration are shown in figures 6 to 7. The respective electrostatic force values on the electrodes are given in table II.

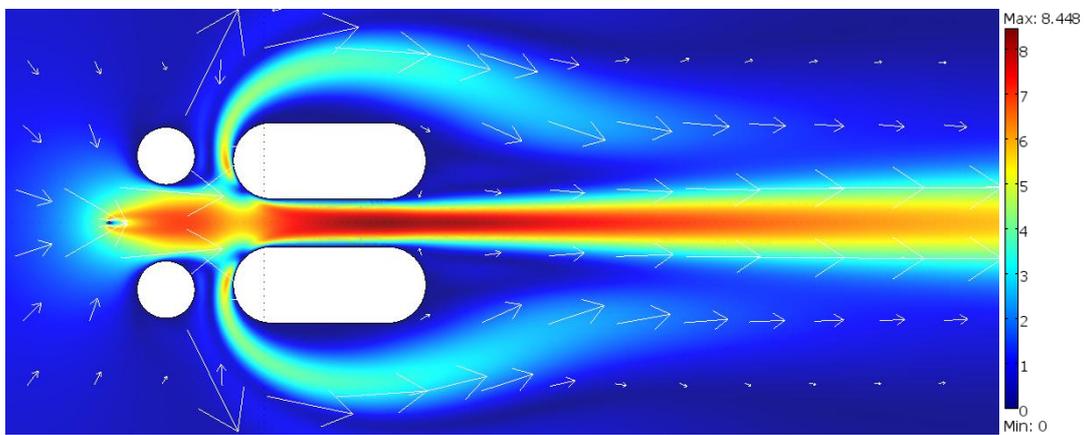

**Figure 3.** Nitrogen velocity as surface map with units in m/s with proportional vector arrows.

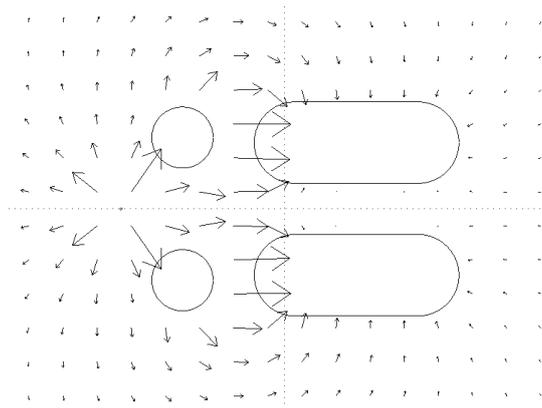

**Figure 4.** Distribution of the electric field vectors (arrows) in the considered structure when the corona discharge is not functioning.

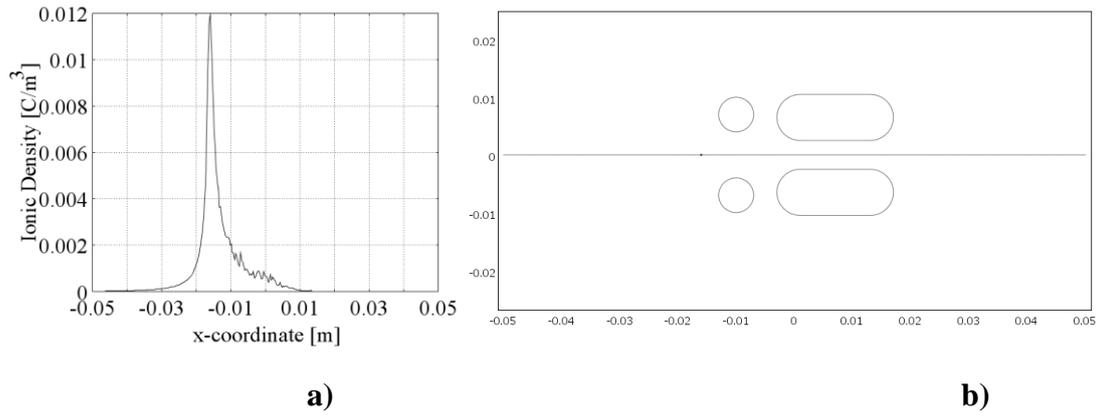

**Figure 5. a)** Ionic density distribution [C/m$^3$] from (-0.05 m, 0 m) to (0.05 m, 0 m), **b)** horizontal line showing where the ion density is taken (units in meters).

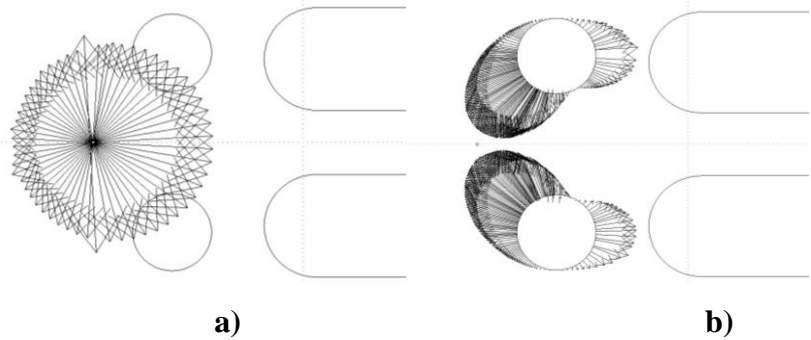

**Figure 6.** Electrostatic force on **a)** the corona wire, and **b)** the ground electrode.

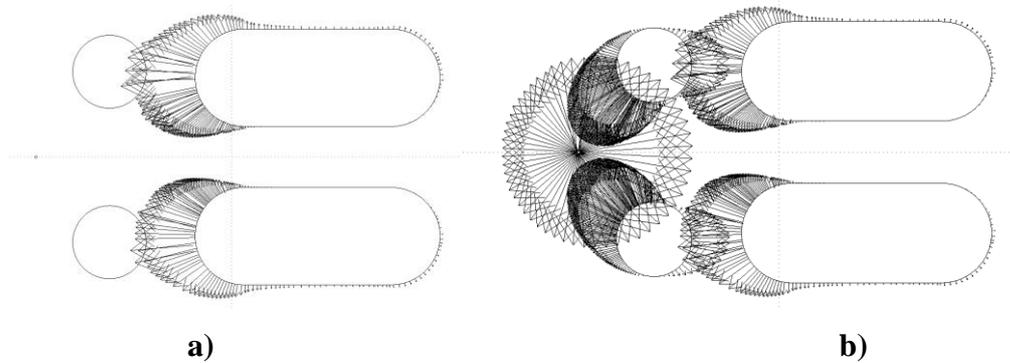

**Figure 7. a)** Electrostatic force on the cathode, **b)** True relative magnitude of the electrostatic force between all electrodes.

**Table II. Electrostatic forces along the x-axis of the capacitor.**

| Geometry | $F_{ex}$ (N/m) |
|---|---|
| Corona wire | 0.0173 |
| Ground upper | -0.0458 |
| Ground lower | -0.0448 |
| Cathode upper | -0.1352 |
| Cathode lower | -0.1341 |
| **Total Force** | -0.3426 |

## IV. CONCLUSION

Our theoretical result, of 0.3426 N/m, for the electrostatic force that acts on the whole capacitor structure is only 2% smaller than the Colas team experimental results[1,2] of 0.35 N/m for the measured thrust per unit electrode length transmitted to the gas. This provides an experimental confirmation and validation to the results obtained in our approximate model in nitrogen. The electrostatic force that acts on the "capacitor" structure is equal and opposite to the force that this structure applies to the surrounding ions and gas.[23,24] This method to theoretically calculate the applied force on the gas from the electrostatic forces that act on the electrode structure is both new and innovative.

This configuration generates a considerable ion wind with top velocity at 8.448 m/s, exactly within the range of the 8-9 m/s measured experimentally in air.[1,2] We have managed to verify theoretically, within our approximation in nitrogen, the experimental measurements performed by Dorian and his colleagues in air.

Figure 4 allows us to see the high electric fields from the corona wire towards the ground electrode responsible for the corona discharge and subsequent nitrogen ionization. Between the ground electrodes and the cathodes we can observe that the electric fields are mostly horizontal as expected, allowing us to confirm the presence of the extra ion acceleration channel responsible for the observed higher velocities in this setup.[2] The ion density goes to zero shortly after the beginning of the cathodes indicating that they strongly attract and neutralize the remaining existing positive ions (figure 5).

The thrust per unit electrode length developed by this setup (0.35 N/m) is slightly higher than the highest value observed (0.327 N/m) in a different setup[23] studied previously where we only have a corona wire (radius of 25 μm) 3 cm above a parabolic cathode with 6 cm of width and 2 cm in height. However the maximum wind velocity developed by this last setup was only of 3.915 m/s compared to the 8.448 m/s of the current setup. This difference may be ascribed to the fact that in the current setup we have a stronger acceleration zone and the gas is forced through a 5 mm channel between the cathodes, where in the former setup the gas flows without constraints above a parabolic surface 6 cm in width. The current setup manages to transmit or channel the electrostatic force on the ions into a smaller volume and therefore the attained neutral wind velocity is much higher. Therefore geometrical considerations and dimensions have a high impact on the performance of these devices, and this will be the focus of future studies on this subject.

## ACKNOWLEDGEMENTS

The author gratefully thanks to Mário Lino da Silva for the permission to use his computer with 32 gigabytes of RAM and two quad-core processors, without which this work would not have been possible.